\documentclass[reprint,eqsecnum,floats,aps,amsmath,amssymb,nofootinbib,prd,onecolumn, showpacs]{revtex4-1}

\usepackage{graphicx}
\usepackage{amsmath,amssymb}
\usepackage{hyperref}
\usepackage{graphicx}
\usepackage{subfigure}
\usepackage{arydshln}
\usepackage{inputenc}
\usepackage{graphicx}
\usepackage{amsmath,amssymb}
\usepackage{hyperref}

\begin{document}

\title{Quantization ambiguities and the robustness of effective descriptions of primordial perturbations in hybrid Loop Quantum Cosmology}

\author{Beatriz Elizaga Navascu\'es}
\email{w.iac20060@kurenai.waseda.jp}
\affiliation{JSPS International Research Fellow, Department of Physics, Waseda University, 3-4-1 Okubo, Shinjuku-ku, 169-8555 Tokyo, Japan}
\author{Guillermo  A. Mena Marug\'an}
\email{mena@iem.cfmac.csic.es}
\affiliation{Instituto de Estructura de la Materia, IEM-CSIC, Serrano 121, 28006 Madrid, Spain}

\begin{abstract}
We study the imprint that certain quantization ambiguities may leave in effective regimes of the hybrid loop quantum description of cosmological perturbations. More specifically, in the case of scalar perturbations we investigate how to reconstruct the Mukhanov-Sasaki field in the effective regime of Loop Quantum Cosmology, taking as starting point for the quantization a canonical formulation in terms of other perturbative gauge invariants that possess different dynamics. This formulation of the quantum theory, in terms of variables other than the Mukhanov-Sasaki ones, is crucial to arrive at a quantum Hamiltonian with a good behavior, elluding the problems with ill defined Hamiltonian operators typical of quantum field theories. In the reconstruction of the Mukhanov-Sasaki field, we ask that the effective Mukhanov-Sasaki equations adopt a similar form and display the same Hamiltonian structure as the classical ones, a property that has been widely assumed in Loop Quantum Cosmology studies over the last decade. This condition actually restricts the freedom inherent to certain quantization ambiguities. Once these ambiguities are removed, the reconstruction of the Mukhanov-Sasaki field naturally identifies a set of positive-frequency solutions to the effective equations, and hence a choice of initial conditions for the perturbations. Our analysis constitutes an important and necessary test of the robustness of standard effective descriptions in Loop Quantum Cosmology, along with their observational predictions on the primordial power spectrum, taking into account that they should be the consequence of a more fundamental quantum theory with a well-defined Hamiltonian, in the spirit of Dirac's long-standing ideas.
\end{abstract}


\maketitle

\section{Introduction}

Effective models are frequently employed to describe (typically low-energy) regimes of many physical processes, that are presumed to have their ultimate theoretical origin in a fundamental quantum theory \cite{eff1,eff2,eff3,eff4,eff5}. This is often the case even if the concrete details about the full underlying quantum system, and its states, are not completely understood. From a theoretical perspective, however, it is important to keep in mind that the details of these effective descriptions generally depend on the freedom that is inherent to the quantization process and on the specific quantum states of the corresponding fundamental theory under consideration. In fact, if this underlying theory is truly fundamental, one typically expects that (many of) the ambiguities that could arise in its quantum description should actually be resolved to guarantee a physically well-behaved and solid structure. For example, a common requirement is that physically important quantities, such as the Hamiltonian that generates the dynamics of the system (or constrains it), are represented in the quantum theory by operators that display a good behavior. Nonetheless, even after the quantum theory is provided with satisfactory bases, there often remains some additional freedom to end its construction, and the way in which this freedom is fixed can strongly affect the effective regimes of the theory.

When discussing ambiguities from quantization choices in the process of building up a theory, we can roughly distinguish two levels on which they can appear in the context of a canonical quantization program. The first one is the quantum representation chosen for the Poisson algebra of the set of elementary canonical variables that describe the system. The second one, that is our subject of study in this paper, concerns the quantization prescriptions adopted for the representation of certain non-linear functions of those elementary variables. Trying and fixing this last freedom is of great relevance, as it is common that non-linear functions of this type play an important physical role and are needed to extract predictions. A especially important case is the Hamiltonian of the theory. The form and properties of its quantum representation as an operator depend on how one fixes the two types of ambiguities that we have commented. Typically, the first of them, concerning the representation of the canonical algebra, can be settled so as to allow for a feasible definition of this Hamiltonian operator. Once this is done, certain effective regimes of physical interest may only be attainable (if any) if one prescribes a specific construction of the non-linear operators that appear in this Hamiltonian.

Here, we focus our attention on a particular theoretical framework where these issues can affect the effective regimes: the loop quantization of a primordial homogeneous and isotropic flat cosmology with inhomogeneous fluctuations. Namely, we consider the application of Loop Quantum Cosmology (LQC) \cite{lqc1,lqc2,abl,ap} techniques to the representation of a homogeneous and isotropic, inflationary flat cosmology, that is supplemented with small metric and matter (anisotropies and) inhomogeneities, regarded as perturbations. LQC is an approach for the quantum description of homogeneous cosmological spacetimes which relies on techniques borrowed from the non-perturbative canonical theory of Loop Quantum Gravity \cite{lqg}. One of its results is a quantum mechanism capable to avoid the Big-Bang singularity by means of a bounce \cite{mmo,APS,taveras}. 

Over the last decade, several approaches have been proposed for the loop quantum description of what are believed to be the primeval stages of our Universe \cite{Bojo1,CLB,Bojo2,hyb-pert1,hyb-pert2,hyb-pert3,hyb-pert4,hyb-pert-eff,hybr-ten,hybr-pred,univ,AAN1,AAN2,AAN3,Ivan,AshNe,Wang1,Wang2,Edward,Edward2,alesci1,alesci2}. Among these approaches, for instance, the one that is commonly called the effective deformed constraint algebra approach \cite{Bojo1,CLB,Bojo2} finds its theoretical motivation precisely on assuming certain types of effective LQC corrections for the homogeneous background, and then imposing that these corrections introduce no anomalies in the constraint algebra. Nonetheless, the effective field equations that result for the cosmological perturbations with this strategy lead to a loss of hyperbolicity in high curvature regimes, and have been argued to produce a primordial power spectrum with enhanced short-scale correlations \cite{Grenoble1,Grenoble2}. Our interest here is rather put on another approach to LQC, the so-called hybrid approach \cite{hyb-pert1,hyb-pert2,hyb-pert3,hyb-pert4,hyb-pert-eff,hybr-ten,hybr-pred,hybr-review}. The corresponding effective description of the cosmological perturbations leads to very different physical results, compatible with the measurements of the Cosmic Microwave Background (CMB) observations \cite{planck,planck-inf,planck19}. This approach rests on a canonical quantization of the full cosmological system with perturbations, truncated at quadratic perturbative order in the Einstein-Hilbert action. In this context, an LQC type of representation is adopted for the homogeneous cosmological sector, while a more conventional Fock quantization is employed to represent the perturbative gauge invariants (e.g. the Mukhanov-Sasaki field for scalar perturbations \cite{MukhanovSasaki,sasaki}). The combination of these different representation techniques is used to construct the operator counterpart of the only constraint of the system that is non-trivial to solve quantum mechanically, namely a global Hamiltonian constraint, that is then imposed \`a la Dirac \cite{Dirac} in order to find physical states \cite{hybrid1,hybrid2}. 

In the hybrid quantization, the representation ambiguities that can affect the effective regimes (and may appear on two different levels as commented above) arise as follows. On the one hand, there is ambiguity in splitting the degrees of freedom between the homogeneous background and the gauge invariant perturbations. These can always be mixed and redefined by means of classical canonical transformations. In particular, since qualitatively different quantum representations are adopted for each sector, any of these splittings can lead to an inequivalent quantum theory \cite{fermiback,msdiag}. Furthermore, as it is common in field theories, there exists ambiguity in the infinite possible choices of a Fock representation for the (gauge invariant) perturbative degrees of freedom. On the other hand, even if one fixes a specific set of canonical variables and a quantum description of them, there is an inherent freedom in constructing non-linear operators from these variables to represent relevant quantities of the system such as the (global) Hamiltonian constraint, which plays a fundamental role in the theory. Some of the quantization prescriptions involved in this last step can have direct consequences even on the effective behavior of such non-linear quantities \cite{mass}. 

Recent investigations have tackled the issue of fixing the ambiguities about the characterization of the perturbive sector and its Fock vacuum, with an eye to attaining well-behaved dynamical properties in the quantum system \cite{msdiag,uniquenessflat,uniquenessrep,fmov,AGvacio1,AGvacio2,no,RMJ}. In particular, a widely known issue of describing the perturbative sector in terms of the Mukhanov-Sasaki field is that its Hamiltonian (and consequently the energy-momentum tensor) is not a well-defined operator on Fock space \cite{AAN1,msdiag}. To avoid the appearance of divergences in its definition, renormalization is necessary. This is a rather unsatisfactory feature in a theory of quantum cosmology that ought to transcend the range of applicability of standard quantum field theory, and avoid its issues traditionally linked to the classicality of spacetime, as has long been argued since Dirac's times. Fortunately, when both the spacetime background and the perturbations are treated as quantum dynamical entities, it has been shown that the aforementioned splitting between them allows one to introduce perturbative gauge invariants with a quantum Hamiltonian that can be perfectly defined on Fock space without regularization \cite{msdiag}. Furthermore, these perturbative variables and their Fock quantization can be chosen so that their Hamiltonian is free of field interactions that would create and destroy infinite types of particles pairs, at least order by order in the ultraviolet regime of short wavelengths. The resulting Hamiltonian, described in terms of annihilation and creation operators for the gauge invariant perturbations, then becomes diagonal. This description of the perturbative sector of phase space, where the dynamics of the gauge invariant variables is ruled by a distinguished set of positive frequencies in the ultraviolet regime, is clearly preferred if the view is put on the construction of a physically satisfactory fundamental theory.

Even though the ambiguities surrounding the choice of canonical variables for the perturbations, and their Fock representation, can be fixed by adhering to this condition of Hamiltonian diagonalization, there remains freedom in the construction of certain non-linear operators in the Hamiltonian. Namely, one still has to prescribe an operator representation for the frequencies of the perturbations, which are non-linear functions of the cosmological background. The aim of this work is to investigate the consequences of these quantization ambiguities on the effective regimes of hybrid LQC, and in particular whether they can compromise the effective equations of the Mukhanov-Sasaki field that have been generally assumed so far. More specifically, in most LQC analyses performed up to date, the dynamics of the Mukhanov-Sasaki field is generated by equations that display a similar behavior as in General Relativity, in what respects its local properties and Hamiltonian structure, but with a different background-dependent mass. Equations of this type have been employed over the last years to extract several observational predictions about the LQC phenomenology on the temperature (and polarization) anisotropies of the CMB \cite{univ,hybr-pred,Ivan,AGvacio2,AshNe,no}. To discuss whether they are a robust consequence of a well-behaved underlying theory is therefore of great importance.

Starting from such a quantum theory, which describes the perturbative sector in terms of variables that evolve diagonally with the aforementioned background-dependent frequencies, the Mukhanov-Sasaki field has to be reconstructed in terms of those variables in effective regimes. The feasibility of this reconstruction, while maintaining the desired form of the effective Mukhanov-Sasaki equations, critically depends on how the background-dependent frequencies behave in the considered effective regime. In this work, we show that an admissible behavior is possible only if one severely restricts the quantum representation of the background functions that define these frequencies. In the effective regime of LQC, where the dynamical relation between the canonical background variables is modified with respect to its classical version, this restriction turns out to affect the quantization prescriptions for the Poisson brackets that provide that relation. As we will see, the resulting reconstruction of the Mukhanov-Sasaki field in this regime is then fixed in terms of its effective background-dependent mass in hybrid LQC. Furthermore, this reconstruction naturally identifies a preferred set of positive-frequency solutions to the effective Mukhanov-Sasaki equations, or equivalently it specifies the initial conditions for the perturbations that are eventually needed in the extraction of predictions.

The outline of this paper is the following. In Sec. II we summarize the canonical formulation of the cosmological system in terms of gauge invariant perturbations, at lowest non-trivial perturbative order. In particular, we  introduce the preferred set of canonical annihilation and creation variables in terms of which the perturbative Hamiltonian becomes diagonal. Furthermore, we outline the main steps for its quantization and the main properties of some associated effective regime, including a discussion of the imprints left in this regime by quantization ambiguities that affect the Poisson brackets between non-linear functions of the background. In Sec. III, starting from the introduced effective regime of the hybrid LQC model, we require that the reconstruction of the Mukhanov-Sasaki field satisfies dynamical equations similar to the classical ones in linearized cosmology. Then, we show that the quantization ambiguities related to Poisson brackets in the Hamiltonian constraint, which are relevant in the effective regime, can be fixed with the previous requirement. We conclude and present our results in Sec. IV. Throughout this paper, we work in Planck units, thus setting $G=c=\hbar=1$.

\section{The model}

Our model can be described by the Einstein-Hilbert action of a Friedmann-Lema\^itre-Robertson-Walker (FLRW) cosmological spacetime in the presence of inhomogeneous and anisotropic perturbations, truncating the action at quadratic perturbative order. The metric of this FLRW spacetime, previous to the introduction of perturbations, can be expressed in terms of a scale factor $a$ and a homogeneous lapse, both of them functions of a global time coordinate $t$. The topology of the flat spatial sections associated with this homogeneous metric, that hereon we call the background metric, is assumed to be compact, isomorphic to the torus $T^3$, and with a compactification length that we set (e.g.) equal to $2\pi$. We use a set of spatial coordinates $\{x^{i}\}$ adapted to the symmetries of this background. To include matter content capable of driving inflation, we minimally couple a homogeneous scalar field subject to a potential. Then, as it is standard in cosmological perturbation theory, we add inhomogeneous modes not only to the metric, but also to this matter field, and truncate their contribution to the total action at quadratic order as well.

The physical information about the inhomogeneities of the cosmological system can be encoded in perturbative gauge invariants, namely fields that do not change under a perturbatively linear spacetime diffeomorphism. In this work, for concreteness, we focus our attention on perturbations of scalar type \cite{mukhanov1,langlois}. For them, one can choose the Mukhanov-Sasaki field and its momentum as the associated gauge invariants. Alternatively, an (in principle) equally valid choice would be any pair of canonical fields that are obtained through a linear canonical transformation of the Mukhanov-Sasaki pair, with coefficients that may depend on the cosmological background. Adopting any of these pairs for the perturbations and developing a canonical formulation for our system, the total Hamiltonian turns out to be a linear combination of constraints that can be written as \cite{hyb-pert4,msdiag}
\begin{align}\label{hamiltonian}
\tilde{N}_{0}[H_{|0}+\tilde{H}_{|2}]+\int_{T^3}d^3\vec{x}\, \tilde{N}^{i}(\vec{x})\mathcal{D}_{i}(\vec{x}) +\int_{T^3}d^3\vec{x}\, K(\vec{x})\mathcal{H}(\vec{x}).
\end{align}
Here, $H_{|0}$ is the function that would correspond to the Hamiltonian constraint of the homogeneous inflationary background if one considered it unperturbed, and $\tilde{N}_{0}$ is a modified homogeneous lapse function that contains quadratic corrections in the perturbations \cite{hyb-pert4}. In addition, $\mathcal{D}_{i}$ and $\mathcal{H}$ are, respectively, Abelianized versions of the linear perturbation of the diffeomorphism and Hamiltonian constraints, with their respective Lagrange multipliers $\tilde{N}^{i}$ and $K$ related with the perturbations of the shift vector and of the lapse function. Besides,
\begin{align}\label{newham}
&\tilde{H}_{|2}=H_{|2}+\frac{1}{2}\sum_{{\vec k} \neq 0}\bigg[2\text{Re}\left({f}^{*}_k\{g_k,H_{|0}\}-{g}^{*}_k\{f_k,H_{|0}\}\right){a}^{*}_{\vec{k}} a_{\vec{k}} + \left({g}^{*}_k\{{f}^{*}_k,H_{|0}\}-{f}^{*}_k\{{g}^{*}_k,H_{|0}\}\right)a_{\vec k}a_{-\vec k} +\text{c.c.} \bigg], \\ \label{h2}&H_{|2}=\frac{1}{2a}\sum_{{\vec k} \neq 0} \left[\big( k^2 + s\big) |v_{\vec{k}}|^2 + |\pi_{v_{\vec{k}}}|^2\right],
\end{align}
where $\vec{k}$ is the wavevector of the corresponding Fourier mode (defined on $T^3$) and can take any non-zero value in $\mathbb{Z}^3$, $k$ is its Euclidean norm, and c.c. stands for the complex conjugate of the previous summand. The relation between Fourier coefficients and functions on $T^3$ is the standard one. For instance, for the Mukhanov-Sasaki field $\mathcal{V}(\vec{x})$, we have $v_{\vec{k}}=\int_{T^3} d^3\vec{x}\, \mathcal{V}(\vec{x}) e^{-i\vec{k}\cdot\vec{x}} /(2\pi)^{3}$ (where $\vec{k}\cdot\vec{x}$ is the Euclidean product of $\vec{k}$ and $\vec{x}$).
We have also defined the following gauge invariant variables in terms of the complex 
Fourier
coefficients $v_{\vec{k}}$ and $\pi_{v_{\vec{k}}}$ of the modes of the Mukhanov-Sasaki field  and its canonical momentum:
\begin{align}\label{anni}
a_{\vec{k}}=f_k v_{\vec{k}} + g_k {\pi}^{*}_{v_{\vec{k}}}, \qquad {a}^{*}_{\vec{k}}={f}^{*}_{k} {v}^{*}_{\vec{k}}+{g}^{*}_{k}\pi_{v_{\vec{k}}},
\end{align}
where the asterisk denotes complex conjugation, and the $k$-dependent factors $f_k$ and $g_k$ satisfy
\begin{align}\label{sympl}
f_k {g}^{*}_{k}-g_k {f}^{*}_{k}=-i .
\end{align}
In principle, we allow that these factors $f_k$ and $g_k$ may depend on the cosmological background. This is why their Poisson brackets with the background Hamiltonian contribution $H_{|0}$ may differ from zero, contributing non-trivially in Eq. \eqref{newham}.
Finally, $s$ is a function of the canonical variables of the cosmological background, that in the following we call the Mukhanov-Sasaki mass. Its explicit expression is not needed for this work and can be found in Refs. \cite{hyb-pert4,mass}. Here, it suffices to say that, when evaluated on classical FLRW trajectories, it provides the usual dynamics in standard linearized cosmology for the Mukhanov-Sasaki field, which is generated by the function $H_{|2}$ \cite{mass,mukhanov1,langlois}.

We have restricted the transformation that defines the collection of gauge invariant variables $a_{\vec{k}}$ and ${a}^{*}_{\vec{k}}$ to be manifestly compatible with the mode symmetries of the Mukhanov-Sasaki Hamiltonian, given by Eq. \eqref{h2}. Their Poisson bracket structure, with only non-vanishing elements equal to $\{a_{\vec{k}},{a}^{*}_{\vec{k}}\}=-i$ owing to condition \eqref{sympl}, has a natural quantum mechanical representation as the commutator algebra of annihilation and creation operators on a Fock space. We henceforth call $a_{\vec{k}}$ and ${a}^{*}_{\vec{k}}$, respectively, annihilation and creation variables.

\subsection{Hamiltonian diagonalization}

If the coefficients $f_k$ and $g_k$ in the definition \eqref{anni} of the annihilation and creation variables are 
just
constants, then those variables simply describe the standard Mukhanov-Sasaki field (and its conjugate momentum). On the other hand, a general dependence of $f_k$ and $g_k$ on the cosmological background permits us to explore different dynamical characterizations of the part of the phase space assigned to the perturbative gauge invariants. Even though the resulting freedom in the choice of annihilation and creation variables translates into the existence of a myriad of inequivalent quantum theories, it also makes it possible to find (hopefully a unique) one with some desirable physical properties. 

This question has been investigated over the last years, and it has been shown that one can guarantee several nice quantum properties for the perturbative Hamiltonian $\tilde{H}_{|2}$ by a convenient selection of the set of annihilation and creation variables \cite{fermiback,msdiag}. In particular, it is well-known that there is no Fock quantization of the Mukhanov-Sasaki contribution $H_{|2}$ that is well-defined on the vacuum state. This is a serious issue, even more since one could expect that a theory that takes into account the quantum dynamical nature of the spacetime should resolve, or at least ameliorate, quantum field theory divergences in the most important operators of the system, as it is the case of those that appear in the constraints. In this context, it has been shown that an appropiate background dependence of $f_k$ and $g_k$ for large $k$ is crucial for a proper definition of the Hamiltonian on Fock space \cite{msdiag}. Moreover, as we explained in the Introduction, it has been argued that one can select a unique preferred set of perturbative variables by demanding that the self-interacting non-diagonal contributions to the Hamiltonian, proportional to $a_{\vec k}a_{-\vec k}$ and its complex conjugate, vanish in Eq. \eqref{newham}, in a way that is completely adapted to its asymptotic structure in the large-$k$ regime. The resulting variables are characterized by a function $h_k=f_k g_k^{-1}$, with strictly negative imaginary part, which satisfies the differential equation \cite{msdiag}
\begin{align}\label{diagallk}
a\{h_k,H_{|0}\}=k^2+s+h_k^{2}
\end{align}
and displays the following asymptotic expansion in the limit $k\rightarrow \infty$:
\begin{align}\label{asymp}
& kh_k^{-1}\sim i\left[1-\frac{1}{2k^2}\sum_{n=0}^{\infty}\left(\frac{-i}{2k}\right)^{n}\gamma_n \right],
\\
\label{recursion}
& \gamma_{n+1}=a\{H_{|0},\gamma_n\}+4s\left[\gamma_{n-1}+\sum_{l=0}^{n-3}\gamma_l \gamma_{n-(l+3)}\right]-\sum_{l=0}^{n-1}\gamma_l \gamma_{n-(l+1)},\qquad \forall n\geq 0, 
\end{align}
with $\gamma_0=s$. This function $h_k$, combined with condition \eqref{sympl}, fixes the coefficients $f_k$ and $g_k$, that define the annihilation and creation variables, except for the phase $F_k$ of $f_k$. Furthermore, if one imposes that the amount of background dynamics that the transformation \eqref{anni} extracts from the phase of the Mukhanov-Sasaki mode coefficients must be minimal, then $F_k$ also becomes fixed as the complex argument of $h_k$, up to a constant \cite{msdiag}. In other words, this last condition sets the phase of $g_k$ equal to an arbitrary constant.

With this choice of annihilation and creation variables, the Hamiltonian $\tilde{H}_{|2}$ adopts the diagonal form \cite{msdiag}
\begin{align}\label{diagh}
\tilde{H}_{|2}=\frac{1}{a}\sum_{\vec k \neq 0}\Omega_k {a}^{*}_{\vec{k}} a_{\vec{k}}, \qquad \Omega_k=-\text{Im}(h_k).
\end{align}
The positive functions $\Omega_k$ of the cosmological background can be regarded as natural frequencies of the gauge invariant perturbations. Henceforth, we refer to this particular description of the perturbations as the  ``variables for Hamiltonian diagonalization''. Since we are interested in the construction of a quantum theory with good physical properties, we focus our attention on this choice of variables in the following.

Let us comment that an analogous treatment could be developed for the tensor perturbations. In fact, their contribution to the total Hamiltonian constraint is almost the same as the Mukhanov-Sasaki one, only differing in the explicit form of the mass $s$ \cite{hybr-ten,mass}. Therefore, all the subsequent discussion can be applied in the tensor case.

\subsection{The quantum theory: Effective regime}

To proceed to the quantization, one has to select a representation for the canonical variables that describe the different sectors of the system (roughly speaking, the background and the perturbations) and use it to construct an operator for the global Hamiltonian constraint\footnote{The linear perturbative constraints restrict the quantum states to depend only on the homogeneous background and the perturbative gauge invariants \cite{hyb-pert4}. Hence, this is the part of phase space that has physical interest for the quantization.}. We recall that this constraint is given by the sum of $H_{|0}$ and $\tilde{H}_{|2}$. The quantum states from which the physical Hilbert space is constructed correspond to the kernel of the (dual) action of this constraint  \cite{Dirac}. Within this general scheme, in this work we focus on the strategy known as hybrid quantum cosmology \cite{hybrid1,hybrid2}. In this approach,  quantum gravity techniques are used to describe the homogeneous background, while a more conventional Fock representation is directly applied to the annihilation and creation variables chosen for the perturbations. In particular, here we adhere to LQC as the quantum-gravity inspired representation for the background. This involves a specific discrete quantization of the background geometry, whereas a standard Schr\"odinger representation on $L^2(\mathbb{R},d\phi)$ is adopted for the inflaton \cite{abl}. 

In the search for physical states of interest in cosmology, one usually focuses the attention on wave profiles such that their variation with respect to the background geometry is affected by the presence of perturbations only in a controllably small way. With this purpose, one typically restricts the analysis to wavefunctions of the form \cite{hyb-pert3,hyb-pert4}
\begin{align}
\Psi(a,\phi,\mathcal{N})=\Gamma(a,\phi)\psi(\mathcal{N},\phi),
\end{align}
where $\mathcal{N}$ is a generic label to denote the mode occupancy numbers in Fock space. The partial wavefunction $\Gamma$ is often chosen close\footnote{In orders of magnitude relative to any suitable perturbative parameter of the system.} to a physical state of the unperturbed FLRW model in LQC. In fact, for the objectives of this work, it is enough to ignore backreaction and consider the limit in which $\Gamma$ is the wave profile of a state consistent with the celebrated effective dynamics of LQC. More specifically, the partial state $\Gamma$ is taken as a physical solution of the inflationary model without perturbations that is highly peaked on effective trajectories of the homogeneous background, which connect a contracting and an expanding branch of the Universe by means of a bouncing mechanism of quantum origin. The bounce occurs at a value of the matter energy density roughly equal to $0.41$ Planck units \cite{abl,APS,taveras}. When the energy density falls significantly below this value, the effective trajectories reduce to the standard FLRW ones. Moreover, these effective trajectories are generated in conformal time by an effective version $H_{|0}^{\text{eff}}$ of the FLRW Hamiltonian $H_{|0}$, via Poisson brackets of the form $a\{.,.\}$. 

Choosing such an effective LQC state for the background part of the wavefunction $\Psi$, one can reasonably expect that there exist situations in which the partial state $\psi$ for the perturbative gauge invariants admits an associated effective regime. Taking into account that the perturbative contribution to the zero-mode of the Hamiltonian constraint is quadratic and diagonal in the annihilation and creation operators, it is rather natural that this effective regime for the perturbations should present a dynamics in conformal time that is generated under Poisson brackets by\footnote{To simplify the notation, in all that follows we denote the effective counterpart of the scale factor of the background cosmology with the same letter $a$ as in the classical formulation, after having fixed $\Gamma$.}
\begin{align}\label{hperteff}
a\tilde{H}_{|2}^{\text{eff}}=\sum_{\vec{k}\neq 0} \left[\Omega_k \right]^{\text{eff}}_{\Gamma}\,a^{*}_{\vec{k}}a_{\vec{k}},
\end{align}
where we use the notation $\left[A\right]^{\text{eff}}_{\Gamma}$ to denote the effective LQC counterpart, associated with the state $\Gamma$, of any non-linear function $A$ of the canonical background variables. So, $\left[A\right]^{\text{eff}}_{\Gamma}$ may correspond to the effective value of a well-defined operator $\hat{A}$ in the hybrid theory through an expectation value on the state $\Gamma$, although our notation allows for other possibilities with a looser correspondance\footnote{For instance, if one starts with a standard densitization of the Hamiltonian constraint, the change to conformal time may lead to a ratio of expectation values, as it happens in previous derivations of the field equations for the scalar perturbations in hybrid LQC \cite{hyb-pert4}.}. 

The quantum representation of the perturbative Hamiltonian clearly involves several ambiguities, even after adopting normal ordering for the annihilation and creation operators, since their coefficients are not linear functions of the canonical variables for the cosmological background. Though one may expect that the consequences of this type of ambiguities should become dilluted on states with a highly peaked effective behavior, in the case of LQC one of them has a strong impact, even in the effective regime that we are considering. It reflects the freedom available in the way to quantize the Poisson brackets with $H_{|0}$ that appear in the expression of $\Omega_k$. Namely,  one can either work out these Poisson brackets before their quantization or, alternatively, represent them as the corresponding operator commutators (multiplied by $-i$). More concretely, given a generic function $F$ of the cosmological background, the following correspondence rules are naturally assumed to hold in the effective regime:
\begin{align}\label{correspondence}
\widehat{\{F,H_{|0}\}}\longrightarrow \{F,H_{|0}\}_{\text{eff}}, \qquad
-i[\hat{F},\hat{H}_{|0}]\longrightarrow \{F,H_{|0}^{\text{eff}}\},
\end{align}
where the hat indicates the LQC-operator representation of the underlying function, and the Poisson brackets $\{.,.\}_{\text{eff}}$ are computed before evaluating the result on the corresponding effective LQC trajectories. Then, one can use any of these two rules to deduce the effective counterpart $[\{F,H_{|0}\}]_{\Gamma}^{\text{eff}}$ of the Poisson bracket of $F$ and the FLRW Hamiltonian $H_{|0}$. However, a distinctive feature of effective LQC is that, in general, one has
\begin{align}
\{F,H_{|0}\}_{\text{eff}}\neq \{F,H_{|0}^{\text{eff}}\}.
\end{align}
Namely, the modifications that change the effective LQC trajectories with respect to the FLRW equations are due to the fact that the relation between the time derivative of the scale factor and its momentum (or the chosen canonical pair of geometric background variables) is changed with respect to the classical one \cite{mass}. Therefore, it is clear that the ambiguities in the representation of Poisson brackets with $H_{|0}$ can leave an important imprint even in the effective regimes of the theory.

An example of a function that involves the type of ambiguity explained above, and that has already been studied in the past, is the Mukhanov-Sasaki mass $s$. In this case, it has been shown that, by working out the Poisson brackets of this function with $H_{|0}$ prior to quantization, one attains more appealing physical properties for the perturbations \cite{mass} (such as the positivity of the mass at the effective LQC bounce). On the other hand, more of these brackets are expected to appear in the contribution $\Omega_k$ associated with the perturbative variables for Hamiltonian diagonalization, because they actually show up in the asymptotic expansion of $\Omega_k$ in the large $k$ limit [see Eqs. \eqref{asymp} and \eqref{recursion}]. From our discussion above, it follows that the dynamical behavior of the effective counterpart $\left[\Omega_k \right]^{\text{eff}}_{\Gamma}$ depends on how one chooses to quantize these brackets. 

\section{Robustness of the Mukhanov-Sasaki equations in LQC}

In this section, we consider the effective regime of the hybrid quantization of the cosmological model that we have presented. Our primary goal will be to show that the ambiguities in the quantization of the Poisson brackets that appear in the frequencies $\Omega_k$ can be fixed in such a way that a satisfactory reconstruction of the Mukhanov-Sasaki field is possible. In fact, as we will see, there is only one way of performing this quantization so that one can consistently arrive at effective Mukhanov-Sasaki equations with a relativistic structure.  In other words, as far as these ambiguities are concerned and within hybrid LQC, there is one way, and only one, to construct a quantum theory with robust physical features such that the dynamics of the Mukhanov-Sasaki field can be exactly of the type that has been assumed so far in the extraction of predictions for observational cosmology \cite{hybr-pred,univ,Ivan,AshNe}. Remarkably, the resulting restrictions on the quantization ambiguities lead to the determination of the effective mass function of the Mukhanov-Sasaki field, in terms of the LQC version of its classical counterpart $s$. Moreover, by fixing $\left[\Omega_k \right]^{\text{eff}}_{\Gamma}$, these restrictions also select a set of positive-frequency solutions for the effective Mukhanov-Sasaki equations, and therefore a vacuum state for the perturbations.

\subsection{Reconstruction of the Mukhanov-Sasaki field}

According to our discussion in Sec. IIB, we take as our starting point that, in hybrid LQC, there exist effective regimes in which the annihilation and creation operators admit counterpart variables that evolve as follows:
\begin{align}\label{effdiag}
a_{\vec{k}}'(\tilde\eta)=-i\tilde{\Omega}_k(\tilde\eta)a_{\vec{k}}(\tilde\eta),\qquad {a}^{*\prime}_{\vec k}(\eta)=i\tilde{\Omega}_k(\tilde\eta){a}^{*}_{\vec{k}}(\tilde\eta),
\end{align}
where $\tilde\eta$ is a generic choice of time parameter in the considered effective description (not necessarily equal to the conformal time $\eta$), the prime denotes derivative with respect to it, and $\tilde{\Omega}_k$ is some real function of this time which is suitably related to $\left[\Omega_k \right]^{\text{eff}}_{\Gamma}$.

In cosmology, one is rather interested in the dynamical equations of the Mukhanov-Sasaki field and of its momentum. As these are no longer elementary canonical fields in the adopted description of the system, we need to reconstruct them from the basic variables in the effective theory. Presumably, this reconstruction would, in turn, have its origins in a definition of the Mukhanov-Sasaki field and its momentum as quantum operators, in terms of the Fock annihilation and creation operators for Hamiltonian diagonalization. Taking into account their classical definition, derived from the inverse of transformation \eqref{anni}, we naturally consider reconstructions of the Mukhanov-Sasaki mode coefficients that are given by the following analog Bogoliubov relations:
\begin{align}\label{tildeMS}
\tilde{v}_{\vec k}=i({\tilde{g}}^{*}_k a_{\vec{k}}-\tilde{g}_{k}{a}^{*}_{-\vec{k}}),\qquad \pi^{*}_{\tilde{v}_{\vec k}} =-i({\tilde{f}}^{*}_k a_{\vec{k}}-\tilde{f}_{k}{a}^{*}_{-\vec{k}}).
\end{align}
Here, $\tilde{f}_k$ and $\tilde{g}_k$ are (yet unspecified) functions of time in the considered effective regimes, that are again subject to
\begin{align}\label{effsympl}
\tilde{f}_k {\tilde{g}}^{*}_{k}-{\tilde{g}}_k {\tilde{f}}^{*}_{k}=-i .
\end{align}
In general, the time dependence of $\tilde{f}_k$ and $\tilde{g}_k$ may be both explicit and/or implicit. The latter type of dependence is here restricted to appear through the dependence on the background variables, that in LQC obey an effective dynamics that differs from the classical FLRW evolution. In any case, we denote the total time derivative of the functions $\tilde{f}_k$ and $\tilde{g}_k$ with the prime symbol. Finally, in the same way as we set the phase of $g_k$ to be constant, we introduce a last restriction on the effective form of the Mukhanov-Sasaki modes by demanding that the phase of the effective analog $\tilde{g}_k$ be time independent.

The evolution equations for our effective Mukhanov-Sasaki variables can be easily computed from Eqs. \eqref{effdiag}-\eqref{effsympl}. Explicitly, one obtains the following time variation:
\begin{align}\label{effdv}
& \tilde{v}_{\vec{k}}'=|\tilde{g}_k|^2\left[\text{Im}(\tilde{h}_k)'+2\tilde{\Omega}_k\text{Re}(\tilde{h}_k)\right]\tilde{v}_{\vec{k}}+2\tilde{\Omega}_k|\tilde{g}_k|^2\pi^{*}_{\tilde{v}_{\vec{k}}},\\
\label{effdpi}
& \pi_{\tilde{v}_{\vec{k}}}^{*\prime}=-|\tilde{g}_k|^2\left[\text{Im}(\tilde{h}_k)'+2\tilde{\Omega}_k\text{Re}(\tilde{h}_k)\right]\pi_{\tilde{v}_{\vec{k}}}^{*}-\left[\text{Re}(\tilde{h}_k)'+2|\tilde{g}_k|^2\big(\text{Re}(\tilde{h}_k)\text{Im}(\tilde{h}_k)'+\tilde{\Omega}_k|\tilde{h}_k|^2\big)\right]\tilde{v}_{\vec{k}}.
\end{align}
For general functions $\tilde{f}_k,\tilde{g}_k$, and $\tilde{\Omega}_k$, these effective equations for the Mukhanov-Sasaki field {\emph{greatly}} differ from the classical ones that hold in the linearized regime of our cosmological system. In particular, the classical equations are cross-diagonal in standard conformal time $\eta$, and take the form
\begin{align}
v'_{\vec{k}}(\eta)=\pi^{*}_{v_{\vec k}}(\eta),\qquad \pi^{*\prime}_{v_{\vec k}}(\eta)=-[k^2+s(\eta)]v_{\vec{k}}(\eta).
\end{align}
Here, we have exceptionally denoted the derivative with respect to $\eta$ with a prime. The structure of these Hamiltonian equations provides particularly nice properties to their solutions. First, their type of cross-diagonality ensures that there is no friction term in the second-order dynamical equation for the Mukhanov-Sasaki field modes. In addition, the specific $k$-dependence in the momentum equation confers locality to the associated field equations, and guarantees their hyperbolicity. In other words, field solutions are uniquely determined [at least locally, for suitable Mukhanov-Sasaki mass function $s(\eta)$] if one gives data at an initial spatial hypersurface. 

In this work, we take a conservative approach and investigate which general conditions are required on our effective scenario if one demands that the (physically and mathematically) well-behaved structure of the classical Mukhanov-Sasaki equations is preserved. Namely, we impose that
\begin{align}\label{effeqsMS}
\tilde{v}'_{\vec{k}}=\pi^{*}_{\tilde{v}_{\vec k}},\qquad \pi^{*\prime}_{\tilde{v}_{\vec k}}=-(k^2+\tilde{s})\tilde{v}_{\vec{k}}.
\end{align}
where $\tilde{s}$ is some (yet unspecified) function of the time $\tilde\eta$, but independent of $k$. It is worth remarking that we are not making a privileged choice of time $\tilde\eta$ by requiring these equations, as this time is not fixed at the moment and we can always effectively redefine it by suitably changing the function $\tilde{\Omega}_k$ [see Eq. \eqref{effdiag}]. We just demand that there exists a time parameter $\tilde{\eta}$ in the effective regime for which the dynamics dictated by Eq. \eqref{effeqsMS} holds. These equations display two important features, that we now analyze in order.
\begin{itemize}
	\item [1)] They are cross-diagonal in the configuration and (complex conjugate) momentum variables. Inspecting the general equations \eqref{effdv} and \eqref{effdpi}, it is clear that this property is satisfied if and only if
	\begin{align}\label{cross-diag}
	\text{Im}(\tilde{h}_k)'+2\tilde{\Omega}_k\text{Re}(\tilde{h}_k)=0.
	\end{align}
	To arrive at the above condition, one has to take into account that relation \eqref{effsympl} implies that $\tilde{g}_k\neq 0$.
	\item [2)] They have a generalized harmonic oscillator structure, in the sense that the time derivative of the configuration variable is equal to the momentum, and the factor that multiplies the configuration variable in the equation for the momentum is given by $-k^2$ plus a $k$-independent function of time. To maintain this structure in Eqs. \eqref{effdv} and \eqref{effdpi}, we need to impose simultaneously that
	\begin{align}\label{omegag}
	&2\tilde{\Omega}_k|\tilde{g}_k|^2=1,
	\\ \label{reim1}
	&\text{Re}(\tilde{h}_k)'+2|\tilde{g}_k|^2\big(\text{Re}(\tilde{h}_k)\text{Im}(\tilde{h}_k)'+\tilde{\Omega}_k|\tilde{h}_k|^2\big)=k^2+\tilde{s}.
	\end{align}
\end{itemize}

The combination of Eqs. \eqref{cross-diag}-\eqref{reim1} with Eq. \eqref{effsympl} results into the following necessary and sufficient condition for our effective Mukhanov-Sasaki variables to obey a dynamics of the standard form \eqref{effeqsMS}:
\begin{align}\label{htilde}
\tilde{\Omega}_k=-\text{Im}(\tilde{h}_k),\qquad \tilde{h}_k'=k^2+\tilde{s}+\tilde{h}_k^2.
\end{align}	
This condition, that involves two relations, imposes quite a strong restriction on the dynamical behavior of the functions $\tilde{\Omega}_k$, since the effective mass function $\tilde{s}$ (though arbitrary) must not depend at all on $k$. On the other hand, we recall that, according to Eqs. \eqref{hperteff} and \eqref{effdiag}, the functions $\tilde{\Omega}_k$ are nothing but effective frequencies of the gauge invariant perturbations, defined with respect to the time $\tilde{\eta}$. Therefore, their form is determined both by the choice of this time and by the quantization ambiguities introduced in Sec. IIB. In the following, we proceed to analyze how all this freedom of choice can be univocally eliminated using the condition \eqref{htilde} that we have just derived for a proper reconstruction of the Mukhanov-Sasaki field.

\subsection{Quantization of Poisson brackets. Positive-frequency solutions}

The Hamiltonian \eqref{hperteff} generates effective equations for the annihilation and creation variables in conformal time $\eta$, which are directly reproduced by Eq. \eqref{effdiag} if we take $\tilde{\Omega}_k=\left[\Omega_k \right]^{\text{eff}}_{\Gamma}$ and $\tilde{\eta}=\eta$. Alternatively, the only other possibility is that we have instead the following relation between the respective times in the evolution equations:
\begin{align}\label{times}
d\tilde{\eta}=\frac{\left[\Omega_k \right]^{\text{eff}}_{\Gamma}}{\tilde{\Omega}_k}d\eta.
\end{align}
We analyze each of these possibilities separately.
\begin{itemize}
\item [1)] Case with time parameter $\tilde{\eta}$ given by the conformal time, and therefore with $\tilde{\Omega}_k=\left[\Omega_k \right]^{\text{eff}}_{\Gamma}$. On classical phase space, we recall that $\Omega_k=-\text{Im}(h_k)$.  Likewise, we have $\tilde{\Omega}_k=-\text{Im}(\tilde{h}_k)$ according to condition \eqref{htilde}. It immediately follows that $\text{Im}(\tilde{h}_k)=\left[\text{Im}(h_k) \right]^{\text{eff}}_{\Gamma}$, so $\left[\text{Im}(h_k) \right]^{\text{eff}}_{\Gamma}$ must satisfy the same second-order differential equation that results from combining the real and imaginary parts of Eq. \eqref{htilde}.  On the other hand, we recall that $h_k$ satisfies the similar equation \eqref{diagallk}. The role of the derivative in this differential equation is played by a (conformal) Poisson bracket with $H_{|0}$. This is the reason why this class of brackets appear in the recursion relation \eqref{recursion} for the asymptotic expansion of $h_k$. Consistency between this and the fact that $\text{Im}(\tilde{h}_k)=\left[\text{Im}(h_k) \right]^{\text{eff}}_{\Gamma}$ requires that
\begin{align}\label{consistency}
[s ]^{\text{eff}}_{\Gamma}-\tilde{s}=\frac{\left[a \{a\{\text{Im}(h_k),H_{|0}\},H_{|0}\}\right]^{\text{eff}}_{\Gamma}-\left[\text{Im}(h_k) \right]^{\text{eff}\prime\prime}_{\Gamma}}{2\left[\text{Im}(h_k) \right]^{\text{eff}}_{\Gamma}} -\frac{3}{4} \frac{\big(\left[a\{\text{Im}(h_k),H_{|0}\}\right]^{\text{eff}}_{\Gamma}\big)^2-(\left[\text{Im}(h_k) \right]^{\text{eff}\prime}_{\Gamma})^{2} }{(\left[\text{Im}(h_k) \right]^{\text{eff}}_{\Gamma})^2}.
\end{align}
The left hand side of this equation is independent of $k$. On the other hand, we recall that the behavior of $h_k$ in the asymptotic limit of large $k$ is fixed by the expansion \eqref{asymp}. In particular, one can check that [see the recursion relation \eqref{recursion}]
\begin{align}\label{imhk}
	\text{Im}(h_k)=-k-\frac{s}{2k}+\mathcal{O}(k^{-3}),
\end{align}
where $\mathcal{O}$ denotes terms of the asymptotic order of its argument, or smaller. Inserting this behavior in Eq. \eqref{consistency}, and looking at the dominant asymptotic order in inverse powers of $k$, one is lead to require that
\begin{align}\label{consistencys}
	\left[s \right]^{\text{eff}\prime\prime}_{\Gamma}=\left[a\{a\{s ,H_{|0}\},H_{|0}\}\right]^{\text{eff}}_{\Gamma}.
\end{align}

Let us consider the commented strategy of evaluating first the Poisson brackets that appear in $\Omega_k$ (outside of the Mukhanov-Sasaki mass $s$), prior to their quantization. This means that the Poisson brackets appearing in Eqs. \eqref{consistency} and \eqref{consistencys} must be written in terms of canonical variables before evaluating them on effective LQC trajectories. The consistency condition \eqref{consistencys} then generally fails in our cosmological model owing to the fact that the usual expression of the Mukhanov-Sasaki mass involves the canonical pair of geometric background variables for which the effective dynamical relation differs from the standard one in General Relativity \cite{mass}. 
	
On the contrary, if we represent these brackets directly as the corresponding operator commutators, we find no tensions at all, since then the effective function $\left[\text{Im}(h_k) \right]^{\text{eff}}_{\Gamma}$ satisfies by construction the same differential equation as $\text{Im}(\tilde{h}_k)$, with $\left[s \right]^{\text{eff}}_{\Gamma}$ playing the role of $\tilde{s}$. Indeed, recalling Eq. \eqref{correspondence} we have that these commutators should correspond to time derivatives in the effective description of the homogeneous background. Moreover, since we have $\text{Im}(\tilde{h}_k)=\left[\text{Im}(h_k) \right]^{\text{eff}}_{\Gamma}$, the requirement that Eq. \eqref{htilde} must hold restricts the mass term $\tilde{s}$ to be equal to $\left[s \right]^{\text{eff}}_{\Gamma}$ in our effective equations for the Mukhanov-Sasaki variables.  All of these results strongly support the quantization of the aforementioned Poisson brackets as commutators.
	
\item [2)] Case with time parameter $\tilde{\eta}$ given by Eq. \eqref{times}, different from the conformal time. We know that the conformal time is a global, $k$-independent, time function. Therefore, if we do not want to introduce any modification in the local structure of our effective Mukhanov-Sasaki equations with respect to the classical one in standard cosmology, $\tilde{\eta}$ must also be $k$-independent. Hence, we need to impose
\begin{align}\label{timed}
\frac{\left[\Omega_k \right]^{\text{eff}}_{\Gamma}}{\tilde{\Omega}_k}=D,
\end{align}
where $D$ is some $k$-independent real function. Taking the derivative of this relation with respect to $\tilde{\eta}$ (denoted again with a prime), we get
\begin{align}\label{derivtime}
(\log{D})'=(\log{\left[\Omega_k \right]^{\text{eff}}_{\Gamma}})'-\frac{\text{Im}(\tilde{h}_k)'}{\text{Im}(\tilde{h}_k)},
\end{align}
where we have used the first relation in Eq.  \eqref{htilde}. If we derive again this expression with respect to $\tilde{\eta}$ and employ the differential equation in \eqref{htilde} for $\tilde{h}_{k}$, we obtain
\begin{align}
k^2-\frac{(\left[\text{Im}(h_k) \right]^{\text{eff}}_{\Gamma})^2}{D^2}=\frac{\left[\text{Im}(h_k) \right]^{\text{eff}\prime\prime}_{\Gamma}}{2\left[\text{Im}(h_k) \right]^{\text{eff}}_{\Gamma}}-\frac{3}{4}\left(\frac{\left[\text{Im}(h_k) \right]^{\text{eff}\prime}_{\Gamma}}{\left[\text{Im}(h_k) \right]^{\text{eff}}_{\Gamma}}\right)^2+\frac{D'}{2D}\frac{\left[\text{Im}(h_k) \right]^{\text{eff}\prime}_{\Gamma}}{\left[\text{Im}(h_k) \right]^{\text{eff}}_{\Gamma}}-\tilde{s}+\frac{1}{4}\left(\frac{D'}{D}\right)^2-\frac{D''}{2D},
\end{align}
where we have employed Eqs. \eqref{timed} and \eqref{derivtime} to write the squares of the real and imaginary parts of $\tilde{h}_k$ and, again, used that $\Omega_k=-\text{Im}(h_k)$. If we now recall the first asymptotic contributions \eqref{imhk} to $\text{Im}(h_k)$, we see that the right hand side of this equation is of order one or smaller, for asymptotically large $k$. However, the left hand side is of the order of $k^2$ for $D\neq 1$, something that clearly leads to a contradiction. Notice that the possibility $D=1$ is ruled out in the considered case, in which $\tilde{\eta}$ differs from $\eta$. We then conclude that, regardless of the quantization prescription for the Poisson brackets in $\Omega_k$, the recovery of the desired effective Mukhanov-Sasaki equations requires the evolution parameter $\tilde{\eta}$ to coincide with the conformal time of effective LQC.
\end{itemize}

The results that immediately follow from our discussion are the following. Mukhanov-Sasaki equations of the standard form of Eq. \eqref{effeqsMS} are truly described by the hybrid LQC model obtained by adopting variables for Hamiltonian diagonalization if (i) the time in those equations corresponds to conformal time, and (ii) one quantizes the Poisson brackets that appear in $\Omega_k$ (other than those implicit in $s$) as operator commutators. Furthermore, in this case the mass $\tilde{s}$ in the effective Mukhanov-Sasaki  equations must be precisely the effective LQC value of the Mukhanov-Sasaki mass, namely $\tilde{s}=\left[s \right]^{\text{eff}}_{\Gamma}$. 

Obviously, the reconstructed Mukhanov-Sasaki variables in our effective description are solutions to those equations. Furthermore, the reconstruction \eqref{tildeMS} straightforwardly selects a specific set of positive-frequency solutions. This set is formed by the functions that multiply the constant annihilation coefficients evaluated at an arbitrary initial time in that relation. Concretely, the annihilation variables evolve diagonally with frequency $\tilde{\Omega}_k=-\left[\text{Im}(h_k) \right]^{\text{eff}}_{\Gamma}$ and one has for them
\begin{align}
k\tilde{h}_k^{-1}=i\left[1-\frac{1}{2k^2}\sum_{n=0}^{\infty}\left(\frac{-i}{2k}\right)^{n}\gamma^{\text{eff}}_n \right],
\end{align}
with $\gamma^{\text{eff}}_{0}=\left[s \right]^{\text{eff}}_{\Gamma}$ and
\begin{align}
\gamma^{\text{eff}}_{n+1}=-(\gamma^{\text{eff}}_n)^{\prime}+4 \left[s \right]^{\text{eff}}_{\Gamma} \left[\gamma^{\text{eff}}_{n-1}+\sum_{l=0}^{n-3}\gamma^{\text{eff}}_l \gamma^{\text{eff}}_{n-(l+3)}\right]-\sum_{l=0}^{n-1}\gamma^{\text{eff}}_l \gamma^{\text{eff}}_{n-(l+1)},\qquad \forall n\geq 0.
\end{align}

Some of the analytic features of this set of solutions have been recently studied in the literature \cite{NO-analy}. In fact, these preliminary investigations support the claim that the primordial power spectrum associated with this set possesses appealing properties, such as a non-oscillatory behavior with respect to both time and the Fourier scale $k$. In this sense, the conclusions reached in this paper serve to clarify how certain preferred initial conditions for the gauge invariant perturbations, sustained on the validity of effective arguments, are actually a direct consequence of a judicious choice of representation and physical state in the hybrid quantum theory.

\section{Conclusions}

We have investigated the role that quantization ambiguities, which are present in the hybrid loop quantization of cosmological perturbations, have on the feasibility of a well-behaved effective theory, and how they can be fixed to provide a quantum theory with good properties that indeed guarantees the existence of such an effective theory. A preferred canonical formulation for the perturbed cosmology, with a quantum Hamiltonian that is well defined, must be one in which the scalar perturbative sector is not described in terms of the conventional Mukhanov-Sasaki field and its momentum, but rather by a linear and background-dependent redefinition of these quantities. The Mukhanov-Sasaki gauge invariants are then not treated as the elementary variables to be quantized, and therefore their reconstruction, in the quantum theory and its effective regimes, is subject to ambiguities. We have shown that requiring that their dynamical equations in effective LQC regimes have the same structure as in the classical description allows us to eliminate quantization ambiguities that otherwise would severely compromise the predictability of the formalism.

In more detail, our cosmological system is a homogeneous and isotropic, flat spacetime that is minimally coupled to an inflaton field and that contains metric and matter perturbations. The Hilbert-Einstein action (together with the matter field action) of this system is truncated at the lowest non-trivial order in the perturbations, namely at quadratic order. The resulting total Hamiltonian is a linear combination of constraints. Among them, the only one that is not straigthforward to impose (at the considered truncation order) is a global one given by the zero-mode of the Hamiltonian constraint of the full system, which is the sum of the contribution of the background cosmology and that of the gauge invariant perturbations. Seeking the construction of a quantum formulation of the full system that is free of the renormalization issues that plague conventional quantum field theory on a fixed background, we have described this contribution of the perturbations in terms of a set of annihilation and creation variables such that the considered term adopts a diagonal expression, completely characterized by an specific asymptotic expansion in the sector of very large Fourier scales.

We have then introduced the main steps for the hybrid quantization of the system, and considered effective dynamical equations with a diagonal structure for these annihilation and creation variables. This type of dynamics is motivated from the quantum theory, when one selects certain states with a highly peaked LQC behavior for the background that endows them, in particular, with an effective Hamiltonian evolution. At this point, starting with a generic expression for the definition of the Mukhanov-Sasaki field and its momentum in the effective regime, similar to their classical expression in terms of our elementary annihilation and creation variables, we have analyzed under which conditions the Mukhanov-Sasaki gauge invariants satisfy equations of motion with the same structure and local properties as in the classical linearized dynamics. These conditions can be translated into a non-trivial requirement on the effective behavior of the positive frequencies of the gauge invariant perturbations. If this requirement cannot be properly met, then a satisfactory justification of the considered type of effective Mukhanov-Sasaki equations would be at stake, considering that they must arise from a quantum theory that is fundamental and robust. This would pose a serious problem from a physical perspective, inasmuch as those equations have been amply employed in the LQC literature to extract observational predictions from the Planckian phenomenology of the very early Universe.

The main result of this paper is that these conditions can be satisfied, and therefore the standard form of the effective Mukhanov-Sasaki equations is truly a consequence of a  well-behaved quantum theory, if certain ambiguities that arise in the hybrid quantization are fixed in a specific way. These ambiguities concern the representation of the Poisson brackets between functions of the background that appear in the perturbative contribution to the global Hamiltonian constraint (other than in the Mukhanov-Sasaki mass term). They are important in the considered effective regimes because the difference in LQC trajectories with respect to the FLRW evolution is due to a modified dynamical relation between the elementary canonical variables of the homogeneous geometry compared to the corresponding classical relation. As a consequence, time derivatives of (functions of) the scale factor on the effective trajectories generally yield a different result than their classical counterparts in terms of Poisson brackets with the Hamiltonian of the FLRW geometry, even if these brackets are evaluated on the same effective trajectories after their explicit computation. Imposing then our requirements about the form of the effective Mukhanov-Sasaki equations, we have shown that these brackets must be quantized as operator commutators, so that they directly lead to time derivatives in the effective LQC regime. Remarkably, this necessary condition is sufficient as well for the recovery of standard Mukhanov-Sasaki equations if their effective time-dependent mass is just the effective value of the corresponding mass operator in hybrid LQC. 

Interestingly, after these ambiguities have been removed, the effective relation for the reconstruction of the Mukhanov-Sasaki field turns out to be fixed in terms of the background dependence of the perturbative Hamiltonian, and of the perturbative variables that evolve diagonally with the natural frequencies associated with this Hamiltonian.  In this reconstruction, it is then straightforward to identify a set of positive-frequency solutions for the Mukhanov-Sasaki equations. A preferred set of this kind is of the utmost importance for a robust extraction of predictions about the primordial power spectrum, and our work provides a strong theoretical motivation for the aforementioned choice.

Beyond the effective regimes, our results on the effective dynamics of the gauge invariant perturbations are important in order to investigate the quantum theory in more detail. Indeed, the quantization prescription for Poisson brackets that we have put forward has a direct impact on the properties of the Hamiltonian constraint operator. It is this constraint what ultimately dictates which quantum states are physical and their dynamical behavior. In fact, our analysis is sufficiently general as to allow a potential extension to other effective scenarios for the cosmological background, beyond standard effective LQC, as long as the dynamical relation between the canonical pair of the homogeneous geometry is different in the effective and in the classical descriptions. For instance, some alternative approaches to the loop quantization of the homogeneous background that have received considerable attention recently, based on a regularization procedure for the curvature operator different than the traditional one, maintain this general property in their effective dynamics \cite{DaporLieg,DaLiTomasz,AlejMena,CGM}. We expect that our study can help to remove similar quantization ambiguities, and check their consequences, in the hybrid quantization of their perturbed versions. 

\acknowledgments
This work was partially supported by Project. No. MINECO FIS2017-86497-C2-2-P and Project. No. MICINN PID2020-118159GB-C41 from Spain. B.E.N. acknowledges the financial support from the Standard program of JSPS Postdoctoral Fellowships for Research in Japan.


\begin{thebibliography}{99}

\bibitem{eff1} C.P. Burgess, An Ode to Effective Lagrangians, published in {\it Radiative Corrections: Application of Quantum Field Theory to Phenomenology}, edited by Joan Sola (World Scientific, Singapore, 1999),  471-488.

\bibitem{eff2} S. Weinberg, Effective field theory, past and future, PoS {\bf CD09}, 001 (2009).

\bibitem{eff3} R. Penco, An introduction to effective field theories, arXiv:2006.16285 (2020).

\bibitem{eff4} J.F. Donoghue, The quantum theory of general relativity at low energies, Helv. Phys. Acta {\bf 69}, 269 (1996).

\bibitem{eff5} C.P. Burgess, Quantum gravity in everyday life: General relativity as an effective field theory, Living Rev. Rel. {\bf 7}, 5 (2004).

\bibitem{abl} A. Ashtekar, M. Bojowald, and J. Lewandowski, Mathematical structure of loop quantum cosmology, Adv. Theor. Math. Phys. {\bf 7}, 233 (2003).

\bibitem{lqc1} M. Bojowald, Loop quantum cosmology, Living Rev. Rel. {\bf 11}, 4 (2008).

\bibitem{lqc2} G.A. Mena Marug\'an, A brief introduction to loop quantum cosmology, AIP Conf. Proc. {\bf 1130}, 89 (2009).

\bibitem{ap} A. Ashtekar and P. Singh, Loop quantum cosmology: A status report, Classical Quantum Gravity {\bf 28}, 213001 (2011).

\bibitem{lqg} T. Thiemann, {\it Modern Canonical Quantum General Relativity} (Cambridge University Press, Cambridge, England, 2007).

\bibitem{APS} A. Ashtekar, T. Paw{\l}owski, and P. Singh, Quantum nature of the big bang, Phys. Rev. Lett. {\bf 96}, 141301 (2006).

\bibitem{taveras} V. Taveras, LQC corrections to the Friedmann equations for a universe with a free scalar field, Phys. Rev. D {\bf 78}, 064072 (2008).

\bibitem{mmo} M. Mart\'{\i}n-Benito, G.A. Mena Marug\'an, and J. Olmedo, Further improvements in the understanding of isotropic loop quantum cosmology, Phys. Rev. D {\bf 80}, 104015 (2009).

\bibitem{Bojo1} M. Bojowald, G. Calcagni, and S. Tsujikawa, Observational constraints on loop quantum cosmology, Phys. Rev. Lett. {\bf 107}, 211302 (2011).

\bibitem{CLB} T. Cailleteau, L. Linsefors, and A. Barrau, Anomaly-free perturbations with inverse-volume and holonomy corrections in loop quantum cosmology, Classical Quantum Gravity {\bf 31}, 125011 (2014).

\bibitem{Bojo2} A. Barrau, M. Bojowald, G. Calcagni, J. Grain, and M. Kagan, Anomaly-free cosmological perturbations in effective canonical quantum gravity, JCAP {\bf 05} (2015) 051.

\bibitem{hyb-pert1} M. Fern\'andez-M\'endez, G.A. Mena Marug\'an, and J. Olmedo, Hybrid quantization of an inflationary universe, Phys. Rev. D {\bf 86}, 024003  (2012).

\bibitem{hyb-pert2}  M. Fern\'andez-M\'endez, G.A. Mena Marug\'an, and J. Olmedo, Hybrid quantization of an inflationary model: The flat case, Phys. Rev. D {\bf 88}, 044013 (2013).

\bibitem{hyb-pert-eff} M. Fern\'andez-M\'endez, G.A. Mena Marug\'an, and J. Olmedo, Effective dynamics of scalar perturbations in a flat Friedmann-Robertson-Walker spacetime in loop quantum cosmology, Phys. Rev. D {\bf 89}, 044041 (2014).

\bibitem{hyb-pert3} L. Castell\'o Gomar, M. Fern\'andez-M\'endez, G.A. Mena Marug\'an, and J. Olmedo, Cosmological perturbations in hybrid loop quantum cosmology: Mukhanov--Sasaki variables, Phys. Rev. D {\bf 90}, 064015 (2014).

\bibitem{hyb-pert4} L. Castell\'o Gomar, M. Mart\'{\i}n-Benito, and G.A. Mena Marug\'an, Gauge-invariant perturbations in hybrid quantum cosmology, JCAP {\bf 06} (2015) 045.

\bibitem{hybr-ten} F. Ben\'{\i}tez Mart\'{\i}nez and J. Olmedo, Primordial tensor modes of the early universe, Phys. Rev. D {\bf 93}, 124008 (2016).

\bibitem{hybr-pred} L. Castell\'o Gomar, G.A. Mena Marug\'an, D. Mart\'{\i}n de Blas, and J. Olmedo, Hybrid loop quantum cosmology and predictions for the cosmic microwave background, Phys. Rev. D {\bf 96}, 103528 (2017).

\bibitem{univ}  B. Elizaga Navascu\'es, D. Mart\'{\i}n de Blas, and G.A. Mena Marug\'an, The vacuum state of primordial fluctuations in hybrid loop quantum cosmology,  Universe {\bf 4} (2018) 98. 

\bibitem{AAN3} I. Agullo, A. Ashtekar, and W. Nelson, A quantum gravity extension of the inflationary scenario, Phys. Rev. Lett. {\bf 109}, 251301 (2012).

\bibitem{AAN1}  I. Agullo, A. Ashtekar, and W. Nelson, Extension of the quantum theory of cosmological perturbations to the Planck era, Phys. Rev. D {\bf 87}, 043507 (2013).

\bibitem{AAN2}  I. Agullo, A. Ashtekar, and W. Nelson, The pre-inflationary dynamics of loop quantum cosmology: Confronting quantum gravity with observations, Classical Quantum Gravity {\bf 30}, 085014 (2013).

\bibitem{Ivan} I. Agullo and N.A. Morris, Detailed analysis of the predictions of loop quantum cosmology for the primordial power spectra, Phys. Rev. D {\bf 92}, 124040 (2015).

\bibitem{AshNe} A. Ashtekar, B. Gupt, D. Jeong, and V. Sreenath, Alleviating the tension in CMB using Planck-scale physics, Phys. Rev. Lett. {\bf 125}, 051302 (2020).

\bibitem{Wang1}  B.-F. Li, P. Singh, and A. Wang, Primordial power spectrum from the dressed metric approach in loop cosmologies , Phys. Rev. D {\bf 101}, 086004 (2020).

\bibitem{Wang2}  B.-F. Li, J. Olmedo, P. Singh, and A. Wang, Primordial scalar power spectrum from the hybrid approach in loop cosmologies, Phys. Rev. D {\bf 102}, 126025 (2020). 

\bibitem{Edward} E. Wilson-Ewing, Testing loop quantum cosmology, Comptes Rendus Physique {\bf 18}, 207 (2017).

\bibitem{Edward2}  F. Gerhardt, D. Oriti, and E. Wilson-Ewing, The separate universe framework in group field theory condensate cosmology, Phys. Rev. D {\bf 98}, 066011 (2018). 

\bibitem{alesci1} E. Alesci, A. Barrau, G. Botta, K. Martineau, and G. Stagno, Phenomenology of quantum reduced loop gravity in the isotropic cosmological sector, Phys. Rev. D {\bf 98}, 106022 (2018).

\bibitem{alesci2} J. Olmedo and E. Alesci, Power spectrum of primordial perturbations for an emergent universe in quantum reduced loop gravity, JCAP {\bf 04} (2019) 030.

\bibitem{Grenoble1} B. Bolliet, J. Grain, C. Stahl, L. Linsefors, and A. Barrau, Comparison of primordial tensor power spectra from the deformed algebra and dressed metric approaches in loop quantum cosmology, Phys. Rev. D {\bf 91}, 084035 (2015).

\bibitem{Grenoble2} S. Schander, A. Barrau, B. Bolliet, J. Grain, L. Linsefors, and J. Mielczarek, Primordial scalar power spectrum from the Euclidean Big Bounce, Phys. Rev. D {\bf 93}, 023531 (2016).

\bibitem{hybr-review} B. Elizaga Navascu\'es and G.A. Mena Marug\'an, Hybrid loop quantum cosmology: An overview, Front. Astron. Space Sci. {\bf 8}, 624824 (2021).


\bibitem{planck} P.A.R. Ade {\it et al.} (Planck Collaboration), Planck 2015 results. XIII. Cosmological parameters, A\&A {\bf 594}, A13 (2016).

\bibitem{planck-inf} P.A.R. Ade {\it et al.} (Planck Collaboration), Planck 2015 results. XX. Constraints on inflation, A\&A {\bf 594}, A20 (2016).

\bibitem{planck19} Y. Akrami  {\it et al.} (Planck Collaboration), Planck 2018 results. VII. Isotropy and statistics of the CMB, A\&A {\bf 641}, A7 (2020).

\bibitem{MukhanovSasaki} V. Mukhanov, Quantum theory of gauge-invariant cosmological perturbations, Zh. Eksp. Teor. Fiz. {\bf 94}, 1 (1988) [Sov. Phys. JETP {\bf 67}, 1297 (1988)].

\bibitem{sasaki} M. Sasaki,  Gauge invariant scalar perturbations in the new inflationary universe, Prog. Theor. Phys. {\bf 70}, 394 (1983).

\bibitem{Dirac} P.A.M. Dirac, {\it Lectures on Quantum Mechanics} (Belfer Graduate School of Science, Yeshiva University, New York, 1964).

\bibitem{hybrid1} M. Mart\'{\i}n-Benito, L.J. Garay, and G.A. Mena Marug\'{a}n, Hybrid quantum Gowdy cosmology: Combining loop and Fock quantizations, Phys. Rev. D {\bf 78}, 083516 (2008).

\bibitem{hybrid2} G.A. Mena Marug\'{a}n and M. Mart\'{\i}n-Benito, Hybrid quantum cosmology: Combining loop and Fock quantizations, Int. J. Mod. Phys. A {\bf 24}, 2820 (2009).

\bibitem{fermiback} B. Elizaga Navascu\'es, G.A. Mena Marug\'an, and S. Prado Loy, Backreaction of fermionic perturbations in the Hamiltonian of hybrid loop quantum cosmology, Phys. Rev. D {\bf 98}, 063535 (2018).

\bibitem{msdiag} B. Elizaga Navascu\'es, G.A. Mena Marug\'an, and T. Thiemann, Hamiltonian diagonalization in hybrid quantum cosmology, Classical Quantum Gravity {\bf 36}, 18 (2019).

\bibitem{mass} B. Elizaga Navascu\'es, D. Mart\'{\i}n de Blas, and G.A. Mena Marug\'an, Time-dependent mass of cosmological perturbations in the hybrid and dressed metric approaches to loop quantum cosmology, Phys. Rev. D {\bf 97}, 043523 (2018).

\bibitem{uniquenessflat} L. Castell\'o Gomar, J. Cortez, D. Mart\'{\i}n-de Blas, G.A. Mena Marug\'an, and J.M. Velhinho, Uniqueness of the Fock quantization of scalar fields in spatially flat cosmological spacetimes, JCAP {\bf 11} (2012) 001.

\bibitem{uniquenessrep} J. Cortez, G.A. Mena Marug\'an, J. Olmedo, and J.M. Velhinho, A uniqueness criterion for the Fock quantization of scalar fields with time-dependent mass, Classical Quantum Gravity {\bf 28}, 172001 (2011).

\bibitem{fmov} M. Fern\'andez-M\'endez, G.A. Mena Marug\'an, J. Olmedo, and J.M. Velhinho, Unique Fock quantization of scalar cosmological perturbations, Phys. Rev. D {\bf 85}, 103525 (2012).

\bibitem{no} D. Mart\'{\i}n de Blas and J. Olmedo, Primordial power spectra for scalar perturbations in loop quantum cosmology, JCAP {\bf 06} (2016) 029.

\bibitem{AGvacio1} A. Ashtekar and B. Gupt, Initial conditions for cosmological perturbations, Classical Quantum Gravity {\bf 34}, 035004 (2017).

\bibitem{AGvacio2} A. Ashtekar and B. Gupt, Quantum gravity in the sky: Interplay between fundamental theory and observations, Classical Quantum Gravity {\bf 34}, 014002 (2017).

\bibitem{RMJ} M. Mart\'{\i}n-Benito, R.B. Neves, and J. Olmedo, States of low energy in bouncing inflationary scenarios in loop quantum cosmology, Phys. Rev. D {\bf 103}, 123524 (2021).

\bibitem{mukhanov1} V. Mukhanov, {\it Physical Foundations of Cosmology} (Cambridge University Press, Cambridge, England, 2005).

\bibitem{langlois} D. Langlois, Inflation and cosmological perturbations, Lect. Notes Phys. {\bf 800}, 1 (2010).

\bibitem{NO-analy} B. Elizaga Navascu\'es, G.A. Mena Marug\'an, and S. Prado, Non-oscillating power spectra in loop quantum cosmology, Classical Quantum Gravity {\bf 38}, 035001 (2021).

\bibitem{DaporLieg} A. Dapor and K. Liegener, Cosmological effective Hamiltonian from full loop quantum gravity dynamics, Phys. Lett. B {\bf 785}, 506 (2018).

\bibitem{DaLiTomasz} M. Assanioussi, A. Dapor, K. Liegener, and T. Paw\l{}owski, Emergent de Sitter epoch of the loop quantum cosmos: A detailed analysis, Phys. Rev. D {\bf 100}, 084003 (2019).

\bibitem{AlejMena} A. Garc\'{\i}a-Quismondo and G.A. Mena Marug\'an, The Mart\'{\i}n-Benito--Mena Marug\'an--Olmedo prescription for the Dapor--Liegener model of loop quantum cosmology, Phys. Rev. D {\bf 99}, 083505 (2019).

\bibitem{CGM} L. Castell\'o Gomar, A. Garc\'{\i}a-Quismondo, and G.A. Mena Marug\'an, Primordial perturbations in the Dapor--Liegener model of hybrid loop quantum cosmology, Phys. Rev. D {\bf 102}, 083524 (2020).
\end{thebibliography}
\end{document}